\documentstyle[aps,prd,epsfig]{revtex} 
\begin{document}
\draft 
\title{
The Photoproduction of the $\boldmath b_1(1235)\pi$ System 
}  
\author{ 
G.R.~Blackett\footnote{\em Present address: 
                  The American Institute of Physics, Woodbury, NY}, 
K. Danyo\footnote{\em Present address:  
                  Brookhaven National Laboratory, Upton, NY}, 
T.~Handler, 
M.~Pisharody\footnote{\em Present address:
                  Argonne National Laboratory, Argonne, IL}, 
G.T.~Condo  
}
\address{ 
Department of Physics                   \\ 
University of Tennessee, Knoxville      \\ 
Knoxville, Tennessee 37996-1200         \\ 
}
\date{\today}
\maketitle
\begin{abstract}
We report on a study of the reaction, 
$\gamma p\rightarrow p \pi^+\pi^+\pi^-\pi^-\pi^0$,
at an incident photon energy of 19.3 GeV.  
The most significant feature of this reaction is $\Delta^{++}$ 
production which occurs with a cross section of $0.6\pm 0.1 \mu$b.
An upper limit is set for the cross section for the reaction, 
$\gamma p\rightarrow\Delta^{++}b_1\!^-(1235)$, and a search is made
for resonances decaying to $b_1\!^\pm\pi^\mp$.
\end{abstract}
Although the non-relativistic quark model has had great success in describing
the well-established light quark mesons as $q\overline{q}$ combinations, it is 
not clear why states with gluonic degrees of freedom remain unobserved.
Close \cite{rf-close-1} has suggested that the preponderance of experimental 
data for $q\overline{q}$ states may merely be reflective of the preponderance
of studies executed with hadronic beams.
In a similar vein, Isgur \cite{rf-isgur-1} has remarked that the identification
of non-$q\overline{q}$ states may require the equivalent of a 
``$J^{PC}$ filter." 
Thus some advantage in a search for non-$q\overline{q}$ states may accrue to 
photoproduction reactions which are dominated by diffractive processes 
(Pomeron exchange) and by one-pion exchange (OPE) processes.

Currently a popular model for describing some non-$q\overline{q}$ mesonic 
states is the flux tube model, originally developed by Isgur, Kokoski and 
Paton \cite{rf-isgur-2}, and later by Close and Page \cite{rf-close-2}
and Barnes, Close and Swanson \cite{rf-barnes-1}.
This model can be used to predict masses, widths and decay modes for 
several possible hybrid states with exotic $J^{PC}$ quantum numbers.
Most of the flux tube hybrid states, predicted in reference \cite{rf-isgur-2}
are experimentally inaccessible either because they decay to very broad 
states ($\Gamma \cong 500$ MeV) or to poorly understood final states, such
as $a_1(1260)$ and $h_1(1170)$, which have themselves been difficult to 
isolate and quantify.  
There are two exceptions in reference \cite{rf-isgur-2} which have decay widths
$\le 300$ MeV, both of which decay to $b_1(1235)\pi$ over 70\% of the time.
These states have the patently exotic quantum numbers $J^{PC} = 0^{+-}$ and 
$1^{-+}$.  
In the more recent flux tube model  calculations by Close and 
Page \cite{rf-close-2}, these states acquire additional decay modes and, 
concomitantly, decay widths exceeding 425 MeV. 
However, there do exist two states in these calculations with widths $\le 300$
MeV and predominant $b_1(1235)\pi$ decay modes.
These states have $J^{PC}=2^{-+}$ and $2^{+-}$.
The latter would appear to be a particularly attractive object to search for
in photoproduction since it has exotic quantum numbers, can be photoproduced
diffractively by a {\em P} wave Pomeron, and is predicted to decay to 
$b_1(1235)\pi$ nearly 90\% of the time.

The purpose of this note is to study the photoproduction of the 
$b_1(1235)^\pm\pi^\mp$ system in the reaction, 
$\gamma p\rightarrow p\pi^+\pi^-\omega\rightarrow p\pi^+\pi^+\pi^-\pi^-\pi^0$,
at an incident photon momentum of 19.3 GeV/c.
As far as we are aware, there is but a single investigation into the 
photoproduction of the $b_1\pi$ final state.
In that study \cite{rf-atkinson-1}, it was concluded that there was evidence
for the production of a $b_1\pi$ final state at $\sim 1880$ MeV which could 
be consistent with flux tube model predictions.  
However that experiment did not identify final state protons, in general, so 
that any resulting $5\pi$ spectrum could contain unknown amounts of 
$\Delta^{++}$ contamination.
Strong photoproduction of the $\Delta^{++}$ via one-pion exchange has 
been shown to be a feature of the experiment we are reporting 
here \cite{rf-condo-1}.

Our data come from a large triggered hydrogen-bubble-chamber experiment which
was performed at the Stanford Linear Accelerator Center, utilizing incident 
photons of average energy 19.3 GeV with a full width at half maximum of 1.7 GeV.
The photon beam was generated by backscattering laser photons from the SLAC
30 GeV electron beam.  
The experimental details have been presented in prior publications
\cite{rf-abe-1}.

In the present experiment we will study events of the type, 
$\gamma p\rightarrow p\pi^+\pi^+\pi^-\pi^-\pi^0$.
The events comprising our sample are those five-prong events which did not have
an acceptable kinematic fit to the three-constraint reaction, 
$\gamma p\rightarrow p\pi^+\pi^+\pi^-\pi^-$.
An acceptable fit required a derived photon energy between 16.5 and 21.0 GeV
and had a probability in excess of $10^{-2}\%$.  
After these events were excluded, we determined those five-prong events with a
single $\pi^0$ by the following procedure.
The beam momentum was set to 19.3 GeV/c, and the square of the missing mass 
was calculated.
If this quantity exceeded 0.1 GeV$^2$, the event was rejected.
If the event survived this cut, the missing mass was treated as a ``track" with
its mass set equal to that of the $\pi^0$.
Lastly a zero-constraint calculation of the five-prong final state was made to
determine the incident photon energy.
If this reconstructed energy fell between 16.0 and 21.5 GeV, the event was
accepted.
With this procedure our overall sample of events of the type, 
$\gamma p\rightarrow p\pi^+\pi^+\pi^-\pi^-\pi^0$, 
consists of 2553 events.

\begin{figure}
  \centerline{\epsfig{file=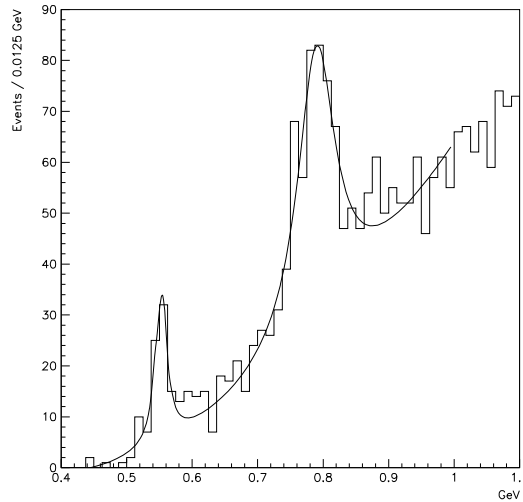,height=3in}}  
  \caption{$\pi^+\pi^-\pi^o$ mass spectrum.} 
  \label{fig1}  
\end{figure}

Since our goal is to study $b_1(1235)^\pm$ photoproduction in the reaction, 
$\gamma p\rightarrow p\pi^+\pi^+\pi^-\pi^-\pi^0$, 
and since the dominant decay of the $b_1(1235)^\pm$ is to $\omega\pi^\pm$, 
we begin by plotting the neutral three pion mass combination 
(Fig.\  \ref{fig1}).
The fit on the histogram gives an $\omega$ mass and width of $790\pm 3$ MeV
and $66\pm 8$ MeV respectively.  
For the $\eta$, we find a mass of $552\pm 2$ MeV and an associated width of 
$17\pm 6$ MeV.
These results seem reasonable considering the nature of the zero-constraint 
fitting process that was adopted. 

\begin{figure}
  \centerline{\epsfig{file=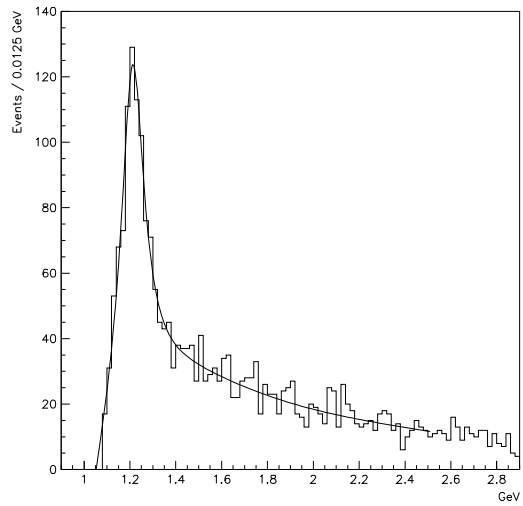,height=3in}} 
  \caption{$p\pi^+$ mass spectrum.}
  \label{fig2}  
\end{figure}

\begin{figure}
  \centerline{\epsfig{file=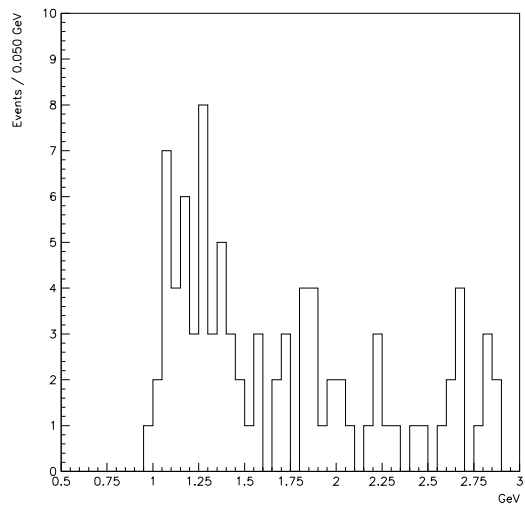,height=3in}}     
  \caption{$\omega\pi^-$ mass spectrum opposite the $\Delta^{++}$.}
  \label{fig3}  
\end{figure}

Before proceeding to the $b_1\pi$ spectrum, we wish to point out that we have
used the present experiment to present evidence for the one-pion exchange
photoproduction reaction \cite{rf-condo-1}, 
$\gamma p\rightarrow a_2(1320)^-\Delta^{++}$, 
with a cross section of $0.45\pm 0.05$ $\mu$b.
We have also observed the reaction,
$\gamma p\rightarrow a_2(1320)^+{n}$, 
with a cross section of $0.29\pm 0.06$ $\mu$b in this 
experiment \cite{rf-condo-2}.
These latter data were found to be consistent with one-pion exchange 
expectations when compared with early low energy ($E_\gamma = 3.5 - 5.0$ GeV)
data \cite{rf-eisenberg-1} as well as with charged $a_2(1320)$ photoproduction
observed at much higher energies \cite{rf-blackett-1} 
($<\!E_\gamma\!> = 110$ GeV).
These data suggest the presence of a $\gamma\pi a_2(1320)$ vertex which had 
been previously quantified by measurements, based on the Primakoff effect, of
the $(\pi\gamma)$ radiative width of the $a_2(1320)$.
This latter experiment \cite{rf-cihangir-1} gave the result, 
$\Gamma(a_2(1320)^\pm\rightarrow\pi^\pm\gamma) = 295\pm 60$ KeV.
In the same experiment \cite{rf-collick-1}, the $(\pi\gamma)$ radiative width
of the $b_1(1235)$ was found to be  
$\Gamma(b_1(1235)\rightarrow\pi^\pm\gamma) = 230\pm 60$ KeV.
With the near equalities of their masses and radiative widths, it is reasonable
to expect comparable one-pion exchange cross sections for 
$\gamma p\rightarrow \Delta^{++}b_1(1235)^-$ and 
$\gamma p\rightarrow \Delta^{++}a_2(1320)^-$. 
Therefore in Fig.\  \ref{fig2} we present the $p\pi^+$ mass spectrum, where a 
large enhancement is observed in the $\Delta^{++}$ mass region.
The fit shown corresponds to a $\Delta^{++}$ cross section of 
$0.6\pm 0.1$ $\mu$b.
Fig.\  \ref{fig3} shows the $\omega\pi^-$ mass spectrum for those events where 
a $\Delta^{++}$ exists.
(We define the $\Delta^{++}$ as $M(p\pi^+)\le 1.4$ GeV and the $\omega$
as $0.73\le M(\omega)\le 0.84$ GeV.)
It is clear that there is, at best, marginal $b_1(1235)^-$ production.
In order to set an upper limit for $\Delta^{++}b_1(1235)^-$ production, we 
assume that all $\omega\pi^-$ events in Fig.\  \ref{fig3} with mass less than 
1.5 GeV comprise our $b_1(1235)^-\Delta^{++}$ signal.
With corrections for our overall detection efficiency ($0.7\pm 0.1$) and for
unseen $b_1(1235)\pi$ decays, the upper limit of the cross section for the 
reaction, 
$\gamma p\rightarrow \Delta^{++}b_1(1235)^-$, 
is $0.04\pm 0.02$ $\mu$b. 
Insofar as the relative photoproduction cross sections for these particles, of
nearly equal mass, can be taken as a measure of their relative $(\pi\gamma)$
decay widths, these data imply that 
$\Gamma(b_1(1235)^\pm\rightarrow\pi^\pm\gamma)\le 0.1\,\Gamma(a_2(1320)^\pm
\rightarrow\pi^\pm\gamma)$. 

This result is clearly inconsistent with the previous measurements of the 
$(\pi\gamma)$ radiative widths for the $a_2(1320)$ and $b_1(1235)$.
However Ishida {\em et.\  al.\  } \cite{rf-ishida-1} have reported a theoretical
value as small as 65 KeV for $\Gamma(b_1(1235)^\pm\rightarrow\pi^\pm\gamma)$.
We should also point out that in the experiment we are reporting on, no 
signal representing $a_1(1260)^-\Delta^{++}$ photoproduction 
\cite{rf-condo-2} could be isolated even though the $(\pi\gamma)$ radiative 
width of the $a_1(1260)$ very likely exceeds that of the $a_2(1320)$
\cite{rf-cihangir-1,rf-zielinski-1,rf-ivanov-1}. 
We are unable to explain the absence of this $J^P=1^+$ state 
in charge-exchange photoproduction.
Another example of this result occurs in the $f_1(1285)\pi$ spectrum where
partial waves representing amplitudes with $J^{PC}=1^{++}$ and $1^{-+}$
were present in $\pi p$ interactions \cite{rf-lee-1}.
This is in contrast to the $f_1(1285)\pi$ spectrum observed from $\gamma p$
interactions where the decay angular distributions suggested the existence
of a $J^{PC}=1^{-+}$ state only \cite{rf-blackett-2}.

\begin{figure}
  \centerline{\epsfig{file=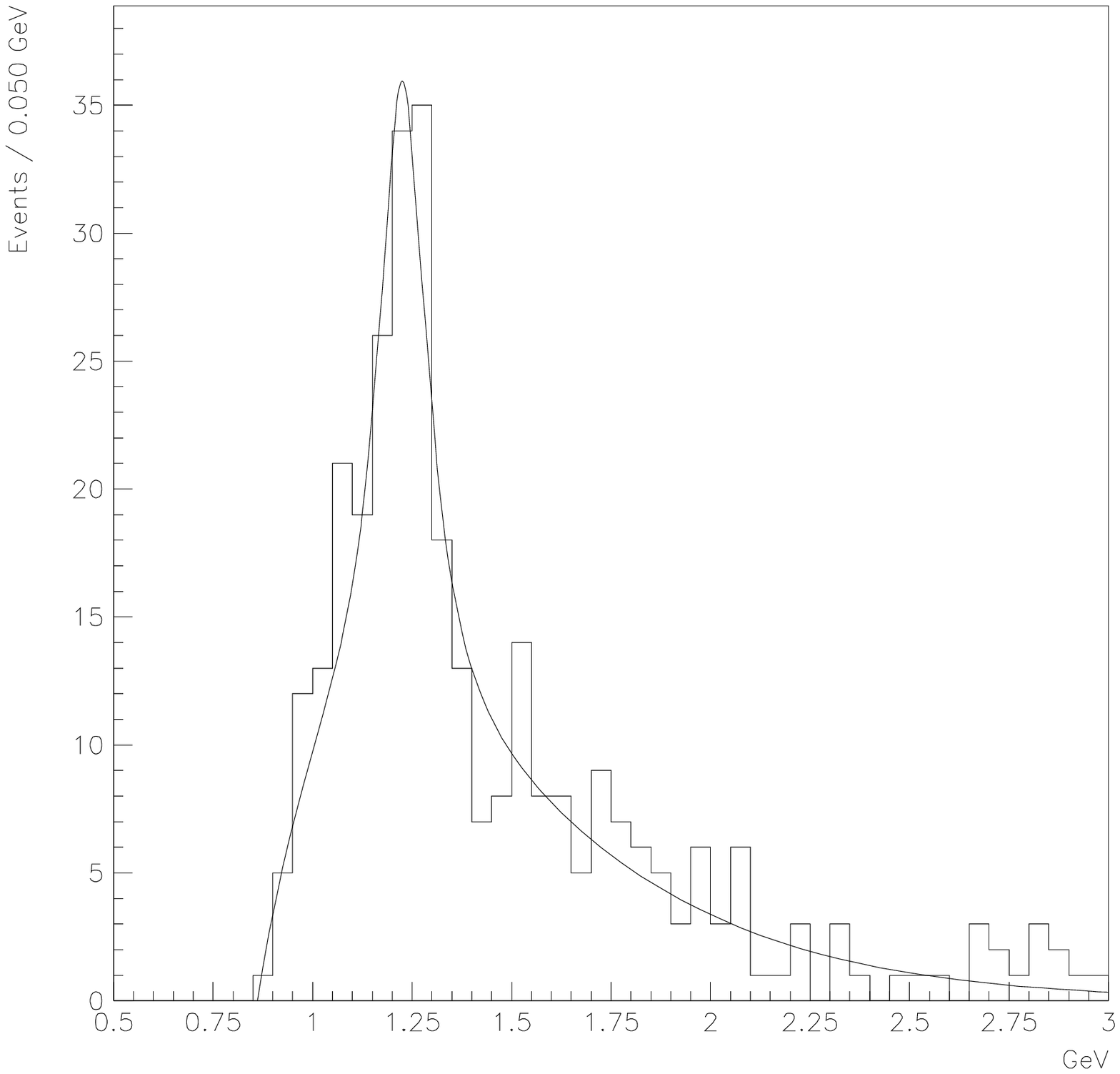,height=3in}     
              \epsfig{file=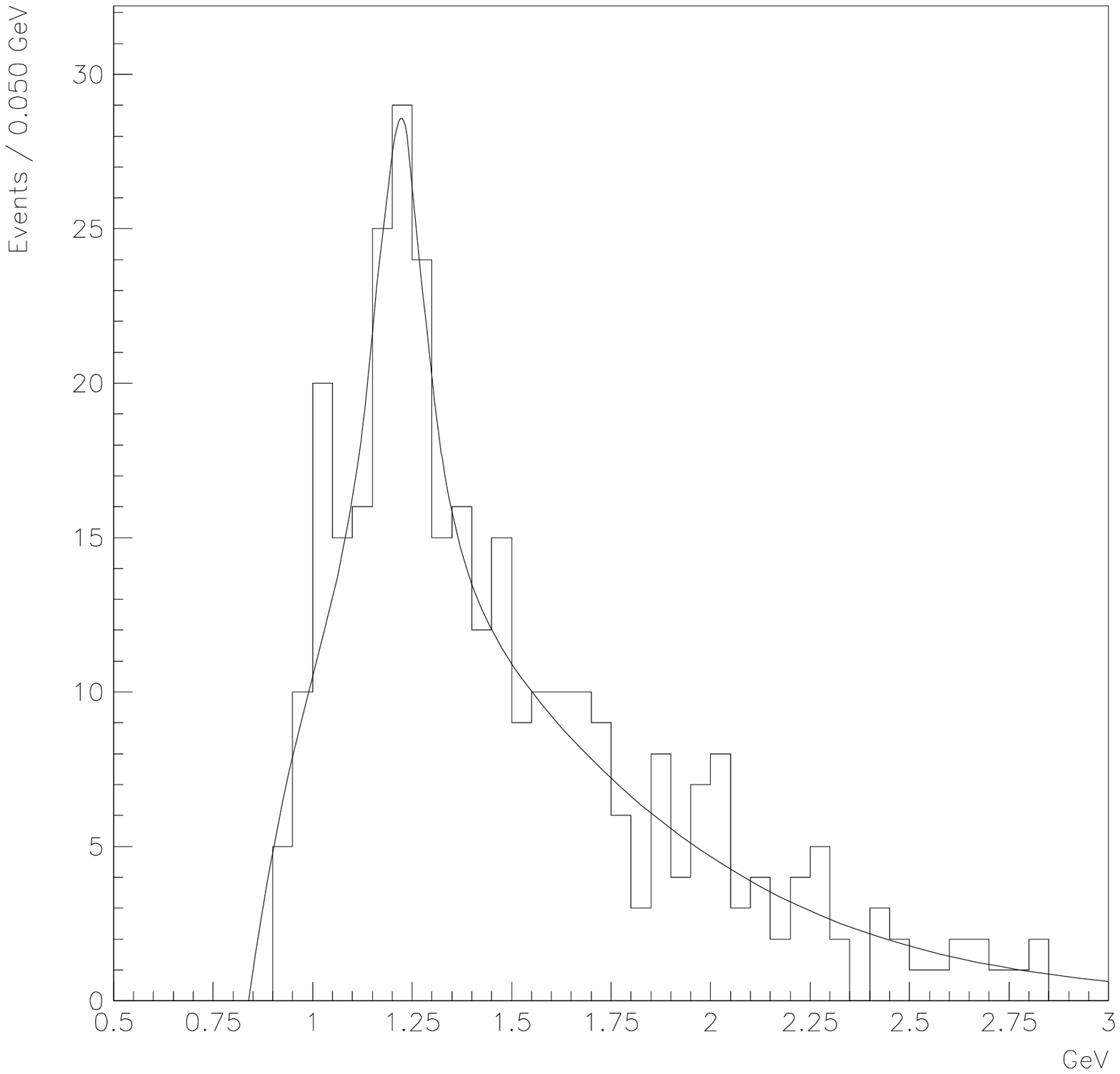,height=3in}}    
  \caption{(a) $\omega\pi^+$ mass spectrum and 
           (b) $\omega\pi^-$ mass spectrum with $\Delta^{++}$ events removed.} 
  \label{fig4}  
\end{figure}

\begin{figure}
  \centerline{\epsfig{file=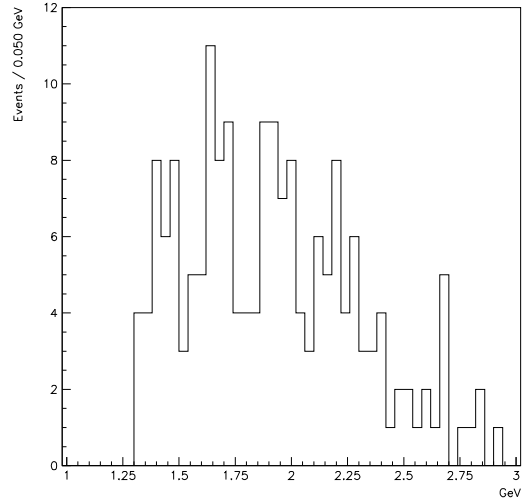,height=3in}}    
  \caption{$b_1(1235)^\pm\pi^\mp$ mass spectrum with $\Delta^{++}$
           events removed.}
  \label{fig5}  
\end{figure}

Because of the large $\Delta^{++}$ signal in Fig.\  \ref{fig2}, the remainder
of this paper will pertain only to the sample of events for which the 
$\Delta^{++}$ has been removed ({\em i.e.}, we accept events only if 
$M(p\pi^+)\ge 1.4$ GeV for both final state pions).
In Figs.\  \ref{fig4}a and \ref{fig4}b we present the $\omega\pi^+$ and
$\omega\pi^-$ mass spectra for this sample.
The only significant enhancement in either spectrum occurs in the mass
region of the $b_1(1235)$ meson.
The fits shown on these plots correspond to the following parameters: \\
$\array{lllllllll}
M(b_1^+) & = & 1.226\pm 0.015 & \rm{GeV} & \hspace{0.2in} &
\Gamma(b_1^+) & = & 0.151 \pm 0.032 & \rm{GeV} \\ 
M(b_1^-) & = & 1.219\pm 0.018 & \rm{GeV} & \hspace{0.2in} &
\Gamma(b_1^-) & = & 0.165 \pm 0.054 & \rm{GeV} \\
\endarray $ 
 \\ 
These values are in tolerable agreement with each other as well as with the 
average values determined by the Particle Data Group \cite{rf-pdg-1}.
If we define the $b_1(1235)$ by the mass cut, 
$1.135\le M(\omega\pi)\le 1.335$ GeV, Fig.\  \ref{fig5} shows the combined
$b_1(1235)^\pm\pi^\mp$ mass spectrum.
While we can identify no significant resonance in this spectrum, it should be
noted that the largest intensities occur in the mass region of the 
$\omega(1670)$ and at $\approx 1900$ MeV.
This result is somewhat similar to that obtained by the Omega Spectrometer
Group \cite{rf-atkinson-1} in the only previous $b_1(1235)\pi$ photoproduction
study.
The failure to observe any $b_1(1235)\pi$ states via photoproduction may not
be too surprising since the quantum numbers of the most likely theoretical 
hybrid candidates are $J^{PC}=1^{-+},\,0^{+-}$ \cite{rf-isgur-2} and
$J^{PC}=2^{-+},\,2^{+-}$ \cite{rf-close-2}.
If we assume that the photoproduction of any of these states proceeds via 
either Pomeron exchange or one-pion exchange, only states with negative charge 
parity and $J\ge 1$ will be present.  
This eliminates all but the (isoscalar) state with $J^{PC}=2^{+-}$.
We are aware of no prior reports of experimental evidence for the existence of
any state with these quantum numbers.
If, on the other hand, the spectrum in Fig.\  \ref{fig5} is interpreted as 
confirming the $b_1(1235)\pi$ enhancement near 1900 MeV, reported by
Atkinson {\em et.\  al.\ } \cite{rf-atkinson-1}, 
the most likely hybrid assignment
for that state would appear to be $J^{PC}=2^{+-}$. 

We wish to thank the many members of the original SLAC BC72,73,75 Collaboration
without whose efforts this work would not exist.
We are also grateful to Ted Barnes for numerous discussions. 


\end{document}